\documentclass[twocolumn,conference]{IEEEtran}
\usepackage[T1]{fontenc}
\usepackage[latin9]{inputenc}
\usepackage{color}
\usepackage{units}
\usepackage{graphicx}
\usepackage[unicode=true,
 bookmarks=true,bookmarksnumbered=true,bookmarksopen=true,bookmarksopenlevel=1,
 breaklinks=false,pdfborder={0 0 0},pdfborderstyle={},backref=false,colorlinks=false]
 {hyperref}
\hypersetup{pdftitle={Your Title},
 pdfauthor={Your Name},
 pdfpagelayout=OneColumn, pdfnewwindow=true, pdfstartview=XYZ, plainpages=false}

\makeatletter

\providecommand{\tabularnewline}{\\}

\usepackage[caption=false,font=footnotesize]{subfig}

\makeatother

\begin{document}
\title{Rate Distortion Theory for Descriptive Statistics}
\author{\IEEEauthorblockN{Peter~Harremo{\"e}s}\IEEEauthorblockA{Niels Brock, Copenhagen Business College\\
Copenhagen, Denmark\\
E-mail: harremoes@ieee.org}}
\maketitle
\begin{abstract}
Rate distortion theory was developed for optimizing lossy compression
of data, but it also has a lot of applications in statistics. In this
paper we will see how rate distortion theory can be used to analyze
a complicated data set involving orientations of early Islamic mosques.
The analysis involves testing, identification of outliers, choice
of compression rate, calculation of optimal reconstruction points,
and assigning ``descriptive confidence regions'' to the reconstruction
points. In this paper the focus will be on the methods,
so the integrity of the data set and the interpretation of the results
will not be discussed.
\end{abstract}

\section{Introduction}

Rate distortion theory was introduced by Shannon as a tool for lossy 
compression. His goal was to compress an information source to a rate
that could be sent through a communication channel with limited capacity.
Applications of ideas from rate distortion theory for testing Goodness-of-Fit
was studied in \cite{Harremoes2008g,Harremoes2019}. The purpose of
the present paper is to demonstrate that rate distortion theory can
be used to solve a variety of statistical problems. The advantage
of these methods is that they are descriptive in nature, which implies
that we only need to make minimal assumptions about how the data were
generated.

To illustrate our method we have analyzed data on the orientation
of early mosques. A database with information on the orientation of
160 early mosques has been compiled by D. Gibson \cite{Gibson2021}.
In this paper we will not discuss the integrity of the data in the
database except that we have excluded the ancient mosque in Aqaba
where it is unclear which wall was the qibla wall.

The data is analyzed using rate distortion theory and the calculations
are done with the R program. There was a rate distortion package for
the R program \cite{Sims2019}, but that package was not maintained.
In addition we need extra features for the type of analysis explained
in the present paper. For this reason we have developed a new package
for R version 4.1.2 for solving rate distortion problems. The software
is developed as a general purpose package and incorporates the
Blahut-Arimoto algorithm and the package is still in development but
the present version is available online \cite{Harremoes2022}. When
fully developed and documented the package will be uploaded to CRAN.
The specific data set has been analyzed using a R worksheet that can
also be downloaded \cite{Harremoes2022a}. 

\section{Historical background}

According to the Islamic traditions Islam was founded in 622 CE, but
the history of early Islam was only written down several hundred years
later. According to these traditions the Muslims were praying facing
Jerusalem until Prophet Muhammad in 624 CE received a revelation
commanding him to face towards the Sacred Mosque (Mashid al-\d{H}ar\={a}m).
Now-a-days Muslims pray facing Mecca in Saudi Arabia where Mashid
al-\d{H}ar\={a}m is located. Inside the Sacred Mosque there is a building
called the Ka'ba, which is at the center of the spiritual life of
the Muslims. 

The Qur'an (Q2:143-144) states that the direction of prayer (qibla)
should be towards Mashid al-\d{H}ar\={a}m, but it is not mention where
Mashid al-\d{H}ar\={a}m is located. The Ka'ba is mentioned as a destiny
of pilgrimage, but it is not mentioned that the Ka'ba is within Mashid
al-\d{H}ar\={a}m. Around 700 CE the Christian author Jacob of Edessa 
wrote that the Arabs pray towards the Ka'ba, but it
does not describe where the Ka'ba was located. Instead the text explains
that the Arabs in Egypt face east, and the Arabs in Kufa pray facing
west \cite{Rignell1979,Hoyland2019}. This does not fit with a Ka'ba
in Mecca, but it is known that there were a number of ka'bas at different
locations in the Arabia \cite[p. 24]{Kalbi1952,Grunebaum2017}.

Early Muslim scholars were aware that many of the oldest mosques did not face 
Mecca in Saudi Arabia. One theory is that the early Muslims
did not know the exact direction towards Mecca. Given their ability
to navigate through the desert this is seems less likely \cite{Gibson2017}.
There are a lot of indications that Islam has its origin in north
western Arabia rather that in the area around Mecca \cite{Crone1987,Khan2021},
and the earliest reference to Mecca outside the Qur'an dates as late
as 743 CE \cite{alTamimi2019}. There is even an old theory that Petra
in Jordan was the birthplace of Prophet Muhammad rather than Mecca
\cite{Hottinger1651}. 

We will use orientations of old mosques to provide information on
the qiblas used during the formative years of Islam. Typically a mosque 
has a long qibla wall with a mihrab (prayer niche) in the middle. 
Muslims face the mihrab and qibla wall when praying. Many early 
mosques have a qibla that appears to be inconsistent with a direction facing Mecca in Saudi Arabia.

The Islamic traditions were written during the Abbassid dynasty
and before that dynasty came into power there had been several civil
wars. In the written accounts the authors openly admit that they have
been selective in their choice of narrative. Since the written accounts
are late and biased towards the ruling Abbasids it is very difficult
to judge which parts of these accounts are historically sound. Here
we will restrict our attention to the period before the Abbasid revolution
in year 750 CE where this Abbasid dynasty came into power. We will
subdivide the period in an early period from 622CE to the reforms
of the Umayyad Caliph 'Abd al-Malik around 700 CE and a late period
from 700 CE to the Abbaside revolution in 750 CE.

\section{Distortion}

In this paper \emph{qibla} is a theory that assigns a certain bearing
to each possible location of a mosque. We will compare the qibla bearing
with the measured orientation of an ancient site. Both the qibla bearing
and the orientation can be given as a number of degrees measured clockwise
from geographical north. Normally the orientation will deviate from
the qibla bearing that was intended by the people who built the mosque.
Four main reasons for this are:
\begin{enumerate}
\item The architect may not have been able to determine the qibla bearing
exactly. 
\item Local obstacles or other practical problems may have influenced the
orientation of the site. 
\item The original structure may now be a ruin or it has been rebuilt so
that the original orientation is difficult to determine.
\item Sometimes it is difficult to measure the orientation as discussed
in \cite{Gibson2017}. 
\end{enumerate}
We have to quantify how much the orientation $o$ deviates from the
qibla bearing $b$ of the site. As distortion measure we use 
\[
\mathrm{versin}\left(o-b\right)=1-\cos\left(o-b\right).
\]
This is the standard method for measuring distortion (also called
dispersion) in directional statistics \cite[Sec. 2.3]{Mardia2000}.
Circular variance can be translated to circular standard deviation
using
\[
\sigma=\frac{360^{\circ}}{2\pi}\left(-2\ln\left(1-V\right)\right)^{\nicefrac{1}{2}}.
\]
The distribution with specified mean direction and specified circular
standard deviation that maximize entropy is a von Mises distribution
\cite{Harremoes2010b}. Thus, using $\mathrm{versin}\left(o-b\right)$
to calculate distortion correspond to using the von Mises distributions
as our basic error model. 

\section{Calculation of the rate distortion function}

As source alphabet we use the sites given by their geographical coordinates
and their bearing. As reconstruction points we use points given by
their coordinates. The distortion of a site and a point is given by
$\mathrm{versin}$ of the difference between the orientation of the
site and bearing from the site to the reconstruction point. We note
that the source alphabet is discrete but the reconstruction alphabet
is continuous. we run the following algorithm in order to calculate
the rate distortion function.
\begin{enumerate}
\item We create a number of random probability vectors over the source alphabet.
These probability vectors are chosen according to a Dirichlet distribution. 
\item For each probability vector over the source alphabet the optimal reconstruction point for these weights is calculated using the Nelder
Mead algorithm.
\item With these reconstruction points we run the Blahut Arimoto algorithm
\cite[Sec. 13.8]{Blahut1972,Cover1991} and get a coupling between
source alphabet and reconstruction alphabet.
\item If one of the reconstruction points has probability close to zero,
it is removed.
\item If the conditional distributions of source points given two different
reconstruction points are close together then one of the reconstruction
points is removed.
\item For each reconstruction point we replace the reconstruction point
by a new reconstruction point that is optimal with respect to the
joint distribution, and go back to to step 3.
\end{enumerate}
The algorithm stops when running 3. to 6. gives an improvement of
the rate below a certain threshold. The Blahut Arimoto algorithm is
iterated until the improvement is below a threshold that is 1/10 of
the threshold used as stopping rule for iterating 3. to 6. 

\section{Test of great circles vs. rhumb lines}

D. Gibson has compared the orientations of the mosques with the geodesic
directions \cite{Gibson2017}, but he has been criticized that calculations
of great circles was not developed in the formative years of Islam
\cite{King2017,King2020}. A method called the Indian Circle is described
in old documents and corresponds to determination of bearings along
rhumb lines \cite{Gibson2021a}. Calculations based on great circles
involve trigonometric formulas and astronomical observation that were
developed by Muslim scholars later than the period that is the focus
of this paper, but in principle a bearing may be determined by other
means than calculations. Using bearings based on rhumb lines has the
advantage that mosques that have the same orientation can be represented
by a very distant reconstruction point.

The question is whether we should base the subsequent calculations
on bearings calculated along great circles taking the curvature of
earth into account or whether we should calculate bearings along rhumb
lines corresponding to a flat earth. Formally, a statistical test
is a binary decision based on data, so we should test great circles
versus rhumb lines. Our decision criteria is simple: we choose the
model that gives the best compression.

\begin{figure}[tbh]
\begin{centering}
\includegraphics[scale=0.3]{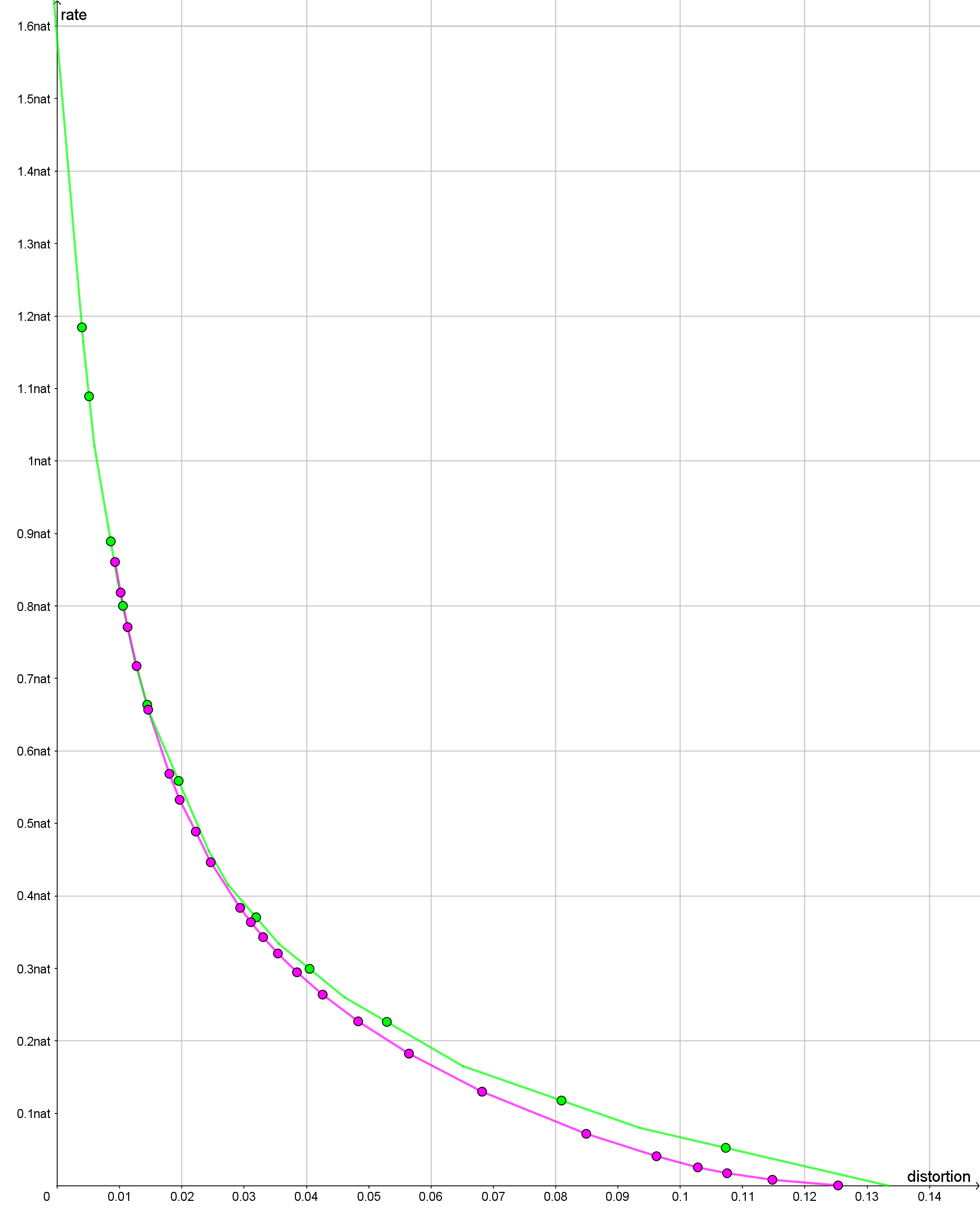}
\par\end{centering}
\caption{\label{fig:FlatVsRound}The green polyline is a lower bound on the
rate distortion function when we use bearings along great circles.
The purple polyline is an upper bound on the rate distortion function
when the bearings are calculated along rhumb lines. The dots are points
where rate and distortion have been calculated.}
\end{figure}

On Figure \ref{fig:FlatVsRound} we compare the rate distortion curve
when the bearings are based on great circles with the rate distortion
curve based on bearings along rhumb lines. For the great circle distortion
we have plotted a lower bound based on the tangents to the rate distortion
curve. For the distortion based on the rhumb lines we have plotted
an upper bound by plotting the cords between achievable rate distortion
pairs. 

As we see the rhumb line distortion gives smaller values of the distortion
than the great circle distortion for rates less than approximately
0.3 nats. For greater rates the two curves are very close together.
The reason that two distortion measures are so similar for high rates
appear to be that the optimal reconstruction points bifurcate and
that the bifurcations tend to blur out the difference between the
two distortion measures. 

Since the bearings based on rhumb lines give a slightly better fit with
data we will use bearings based on rhumb in the rest of this paper. 

\section{Outlier detection for sites before 700 CE}

The rate distortion curve is parametrized by its slope $s.$
For each slope $s$ we get a list of reconstruction points and a joint
distribution of sites and reconstruction points. Both the reconstruction
points and the joint distribution will depend on the rate, but the
individual reconstruction points are quite robust to changes in the
rate. If we increase the rate the reconstruction point will split
into a large number of reconstruction points each with little weight.
If we decrease the rate then reconstruction points will merge together.
It requires a little experimentation to find a rate that gives interesting
result. Here we will use the slope $s=-83$. Further justification
of this value will be given in Section \ref{sec:Determination-of-rate}.

First we will make a rate distortion analysis on the set of the 20
sites that are dated prior to year 700 CE. If we compress with slope
$s=-83$ we get a mean distortion of 0.00395 corresponding to a circular
standard deviation of $5.2^{\circ}.$ At this slope the rate is 0.4840
nats. We get the following reconstruction points. 

\begin{table}[tbh]
\begin{centering}
\begin{tabular}{|c|c|c|c|c|}
\hline 
\textbf{Rec. point} & \textbf{Latitude} & \textbf{Longitude} & \textbf{Weight} & \textbf{Distortion}\tabularnewline
\hline 
\textbf{Pe} & $30.1439^{\circ}$ & $35.4267^{\circ}$ & 77.16 \% & 0.00446\tabularnewline
\hline 
\textbf{\textcolor{blue}{Ma}} & $18.5177^{\circ}$ & $28.7456^{\circ}$ & 15.34 \% & 0.00266\tabularnewline
\hline 
\textbf{\textcolor{red}{SG}} & $-5.4385^{\circ}$ & $-33.4956^{\circ}$ & 7.50 \% & 0.00134\tabularnewline
\hline 
\end{tabular}
\par\end{centering}
\caption{Reconstruction points for $s=-83.$ The labels of the reconstruction
points are based on the interpretations of the clusters given later
in this paper.}
\end{table}

In order to identify these reconstruction points we make a table of
the conditional probabilities of the reconstruction points given the
site.

\begin{table}[tbh]
\begin{centering}
\begin{tabular}{|c|c|c|c|}
\hline 
\textbf{Site} & \textbf{Pe} & \textbf{\textcolor{blue}{Ma}} & \textbf{\textcolor{red}{SG}}\tabularnewline
\hline 
Massawa Mosque & 100.0 & 0.0 & 0.0\tabularnewline
\hline 
Huaisheng Mosque & 86.9 & 0.1 & 13.0\tabularnewline
\hline 
Hama Great Mosque & 91.1 & 8.9 & 0.0\tabularnewline
\hline 
Palmyra Congregational & 92.2 & 7.8 & 0.0\tabularnewline
\hline 
Amr ibn -Al-As & 100.0 & 0.0 & 0.0\tabularnewline
\hline 
\textcolor{red}{Sidi Ghanem} & 0.0 & 0.0 & \textbf{\textcolor{red}{100.0}}\tabularnewline
\hline 
\textcolor{blue}{Graveyard of Sidi 'Ukba} & 0.1 & \textbf{\textcolor{blue}{58.2}} & 41.7\tabularnewline
\hline 
Qasr Humeima & 100.0 & 0.0 & 0.0\tabularnewline
\hline 
\textcolor{blue}{Zawailah} & 0.0 & \textbf{\textcolor{blue}{100.0}} & 0.0\tabularnewline
\hline 
Dome of the Chain & 100.0 & 0.0 & 0.0\tabularnewline
\hline 
Ka'ba & 100.0 & 0.0 & 0.0\tabularnewline
\hline 
Qasr El-Bai'j & 79.2 & 29.6 & 0.0\tabularnewline
\hline 
Um Jimal Later Castellum & 84.7 & 15.3 & 0.0\tabularnewline
\hline 
Kathisma Church & 100.0 & 0.0 & 0.0\tabularnewline
\hline 
Qasr Mushash & 86.1 & 13.9 & 0.0\tabularnewline
\hline 
Seven Sleepers Mosque & 92.1 & 7.9 & 0.0\tabularnewline
\hline 
Husn Umayyad Mosque & 97.4 & 2.6 & 0.0\tabularnewline
\hline 
Zeila Qiblatain Mosque (Rt) & 100.0 & 0.0 & 0.0\tabularnewline
\hline 
Zeila Qiblatain Mosque (Lft) & 100.0 & 0.0 & 0.0\tabularnewline
\hline 
\end{tabular}
\par\end{centering}
\caption{Soft classification of sites older than 700 CE. Probabilities are
given in percent.}
\end{table}

First we observe that the Sidi Ghanem mosque is the only mosque that
has a significant contribution to the reconstruction point SG. The mosque has been
rebuilt many times and it is not clear which wall was the original
qibla wall of this mosque \cite{Gibson2018}. Compared with the rest
of the data set we consider Sidi Ghanem as an outlier and we remove
in from the data set.

If the Sidi Ghanem mosque is removed from the data set Graveyard of
Sidi 'Ukba will get about 99.9 \% probability of reconstruction point
Ma and 0.1 \% probability of reconstruction point Pe. Thus the reconstruction
point Ma essentially only have contributions from the Graveyard of
Sidi 'Ukba and the Zawaila Congretional Mosque. The rest of the sites
only give marginal contributions to this reconstruction point. These two sites and Sidi
Ghanem all lie in Magreb, i.e. North Africa to
west of Egypt. Gibson has classified all sites in his database in
Magreb as having the ``parallel qibla'' \cite{Gibson2017}. Since
there are only these three sites in the Magreb from this early period,
we can only observe that their qiblas are significantly different
from the qiblas from the rest of the sites. Here we will consider
these three sites as outliers of the data set, i.e. they are so untypical
that we will remove them from the data set and analyze the rest of
the data set without them.
\begin{figure}[tbh]
\begin{centering}
\includegraphics[scale=0.3]{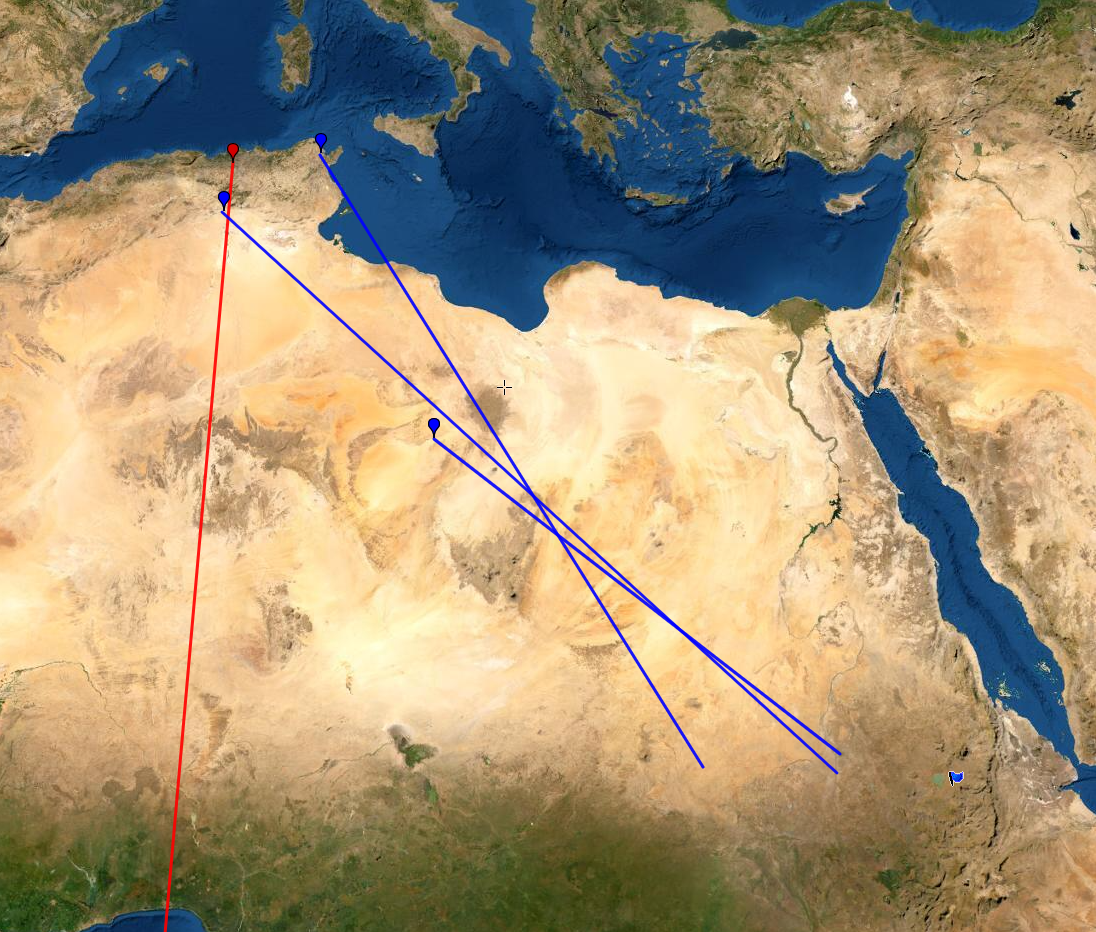}
\par\end{centering}
\caption{\label{fig:Magreb}The red marker and line indicate the Sidi Ghanem
Mosque and its orientation. The blue markers and blue full lines indicate
the locations and bearings Sidi 'Ukba Graveyard and the Zawailah mosque.
The blue square is the optimal reconstruction point of cluster Ma.
The reconstruction point MA is not at the intersection of the blue
lines because there are also some small contributions from sites in
the Levant. The blue marker with the dashed line illustrates the only
mosque in Magreb between 700 CE and 750 CE.}
\end{figure}

\section{\label{sec:Determination-of-rate}Determination of rate and reconstruction
point}

When the outliers have been removed and we use $s=-68$ as the value
of the slope we get a single reconstruction point with coordinates
$30.1286^{\circ}N$ $35.4170^{\circ}E$. The mean distortion for these
early sites is 0.00481 corresponding to a circular standard deviation
of $5.6^{\circ}.$ If the slope is lowered $s\leq-69$ the optimal
reconstruction point starts to bifurcate into a number of optimal
reconstruction points that are located very close to each other as
illustrated in Table \ref{tab:Bifurcation}. This is a strong indication
that compression with $s\leq-69$ leads to compression of the noise rather
than the signal. 
\begin{table}[tbh]
\begin{centering}
\begin{tabular}{|c|c|c|c|c|}
\hline 
\textbf{Rec. point} & \textbf{Latitude} & \textbf{Longitude} & \textbf{Weight} & \textbf{Distortion}\tabularnewline
\hline 
\textbf{Pe1} & $30.1286^{\circ}$ & $35.4170^{\circ}$ & 99.87 \% & 0.00481\tabularnewline
\hline 
\textbf{Pe2} & $30.5953^{\circ}$ & $35.3761^{\circ}$ & 0.13 \% & 0.00498\tabularnewline
\hline 
\end{tabular}
\par\end{centering}
\caption{\label{tab:Bifurcation}Reconstruction points at slope $s=-69$ for sites dated
before 700 CE. Outliers were removed. }
\end{table}

\section{Calculation of a descriptive confidence region}

The use of confidence regions is widely used in statistics, but due
to the complexity of the data and the model we do not have formulas
for calculating such confidence regions. Instead we will use bootstrap
techniques as described in \cite[Sec. 5.2]{Voss2014} to calculate
a region that resembles the well-known notion of a confidence region.
The interpretation is closely related to the notion of cross validation.

The optimal reconstruction point Pe is obtained by minimizing
the mean distortion where each of the 17 sites has weight 1. One may
argue that a large congressional mosque should have larger weight
than a small rural mosque. If a mosque has two qiblas one may ask
if each of the two qiblas should have the same weight as a mosque
with a single qibla. One may also ask if a mosque is rebuilt with
the same qibla should count as one or two. Finally some may question
the dating of some of the mosques. They may argue that some of the mosques
should be removed from the data set. One of the main purposes in natural
sciences for making controlled experiments is to obtain exchange-ability
of the individual results. Exchange-ability implies that all data
points should have the same weight. In humanities we often face the
problem that data is not collected by controlled experiments. Therefore
there is no default reason why all sites should have the same weight.
We will examine what happens if we randomly assign weights
to the sites before we calculate the optimal reconstruction point. 

The random weights are assigned be re-sampling. From the 17 sites
we sample 17 sites with replacement. In such a bootstrap sample only
about 67 \% of the original sites will appear and some of the sites
will appear several times such that the sites in the bootstrap sample
will have multiplicities that sum to 17. This corresponds to assigning
random integer weights to the sites. For this bootstrap sample we
find the optimal reconstruction point where the mean distortion is
calculated with weights of the sites given by the multiplicity specified
by the bootstrap sample. In principle we should go through all possible re-samples.
Then we approximate the distribution of reconstruction points by a
2 dimensional Gaussian distribution. Finally we calculate the ellipse
that contains 95 \% of probability mass of the 2 dimensional Gaussian
distribution and this will be our descriptive confidence region.

If the same procedure is used for a binomial distribution one will
get the formula for calculation of the z-interval of the success probability.

\begin{figure}[tbh]
\begin{centering}
\includegraphics[scale=0.25]{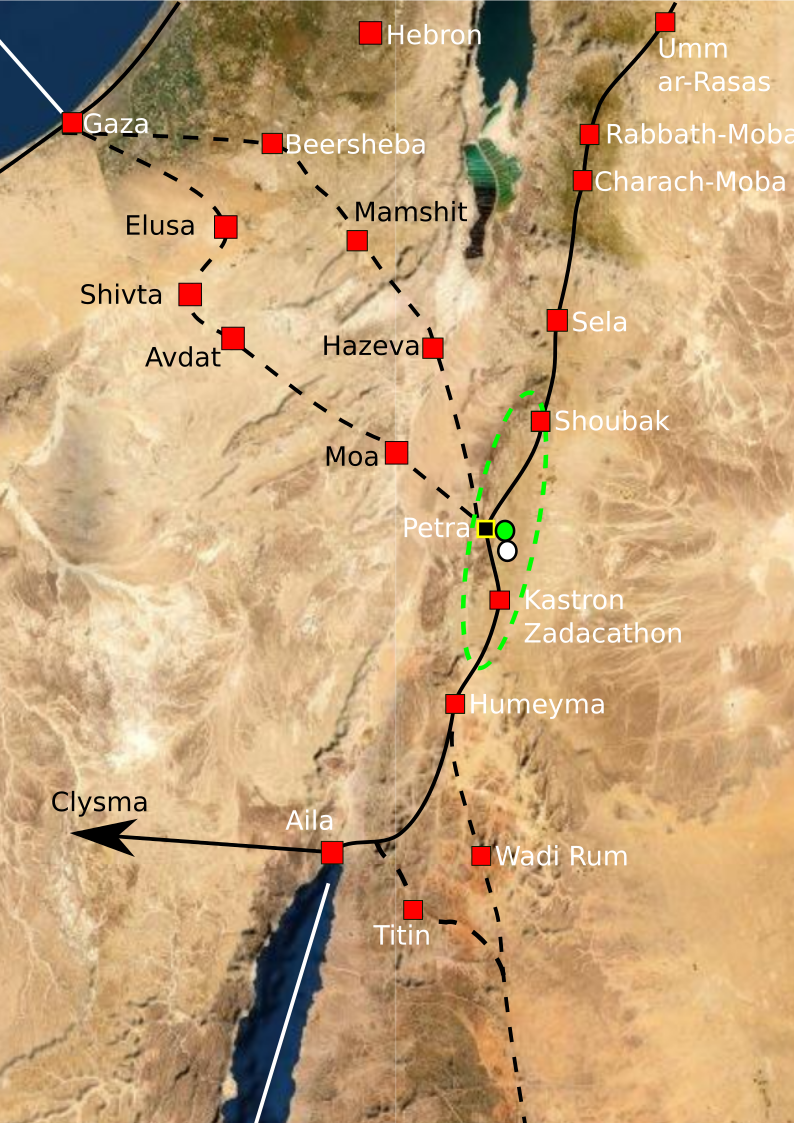}
\par\end{centering}
\caption{\label{fig:Bootstrap}The white dot is the optimal reconstruction
point for cluster Pe based on 17 sites. The confidence region is outlined
by the green ellipse and its center is marked by a green dot.}
\end{figure}

Instead of going through all $17^{17}$ ways of re-sampling we just
randomly take 10000 re-samples and base our calculations on that. Bootstrap re-sampling is implemented by the bootstrap package in the
R program. The resulting descriptive confidence region is depicted
in Figure \ref{fig:Bootstrap}. The descriptive confidence region
is closely confined around the ancient city of Petra. Inside the region
there are no obvious alternative candidates for an early Islamic qibla. 

Around 700 CE Jacob of Edessa stated that the mhaggriiye (Arabs) pray
towards the Ka'ba. Since the optimal reconstruction point is consistent
with the text of Jacob of Edessa, we believe that he refers to the
qibla associated with cluster Pe. There must have existed a Ka'ba
in the confidence region around the reconstruction point Pe before 700 CE. 

As our descriptive confidence method shows the conclusion will
not alter if we remove a few of the sites from the data set or if
we gave the sites slightly different weights. One problem about this
analysis is that it may suffer from selection bias. In this period
the mosques do not have mihrabs so the identification of whether a
site has a qibla may have been effected by how the building is oriented
compared with Petra and Mecca. For this reason we will use the later
mosques to cross validate our temporary conclusion that early Muslims
outside Magreb used Petra as qibla. 

\section{Cross validation using later mosques}

The analysis of mosques from before 700 CE suggests that the early
Muslims were able to determine the qibla with a mean distortion of
0.00481 corresponding to a circular standard deviation of $5.6^{\circ}.$
Now we can compress the data involving all sites before 750 CE using
the same distortion level. A short summery of the results are as follows:
\begin{itemize}
\item All the mosques in the Magreb appear as outliers and are removed from
the data set.
\item The reconstruction point Pe associated with Petra appear again with
approximately the same coordinates and the same confidence region.
\item A reconstruction point Je appears with Jerusalem as qibla. Only a
single mosque at Qasr Tuba is associated with this qibla with high
confidence. A few other mosques could also have had this qibla. 
\item Two new reconstruction points emerge south and south east of Petra.
If the slope is increased to $s\geq-29$ then these two reconstruction
points merge into a single reconstruction point that we will label Ru. 
\end{itemize}
One may conjecture that the early Muslims were, for some unknown reason,
not able to determine the bearing to the reconstruction point Ru with the same precision
as the mosques facing Petra. The result can be seen in Table \ref{tab:ClustersSlope-30}.
With this compression the distortion of the reconstruction point Ru is 0.01329
corresponding to a circular standard deviation of $9.4^{\circ}.$
The mosques associated with the reconstruction point Ru are approximately
the same as the ones that D. Gibson classified as having ``the between
qibla'' \cite{Gibson2017}. According to Dan Gibson's theories ``the
between qibla'' was used because of political tensions between the
ruling Umayyad dynasty and the Abbas family that would later establish
the Abbasid dynasty. According to his theories the mosques that used
``the between qibla'' used the bisector or median between the bearing
to Petra and the bearing to Mecca.

\begin{table}[tbh]
\begin{centering}
\begin{tabular}{|c|c|c|c|c|}
\hline 
\textbf{Rec. point} & \textbf{Latitude} & \textbf{Longitude} & \textbf{Weight} & \textbf{Distortion}\tabularnewline
\hline 
\textbf{Je} & $31.7781^{\circ}$ & $35.2353^{\circ}$ & 2.57 \% & 0.00662\tabularnewline
\hline 
\textbf{Pe} & $30.3289^{\circ}$ & $35.4433^{\circ}$ & 34.01 \% & 0.01134\tabularnewline
\hline 
\textbf{Ru} & $27.6664^{\circ}$ & $36.2188^{\circ}$ & 63.42 \% & 0.01329\tabularnewline
\hline 
\end{tabular}
\par\end{centering}
\caption{\label{tab:ClustersSlope-30}Clusters at slope $s=-29$ with outliers
removed and Petra and Jerusalem fixed.}
\end{table}
\begin{figure}[tbh]
\begin{centering}
\includegraphics[scale=0.28]{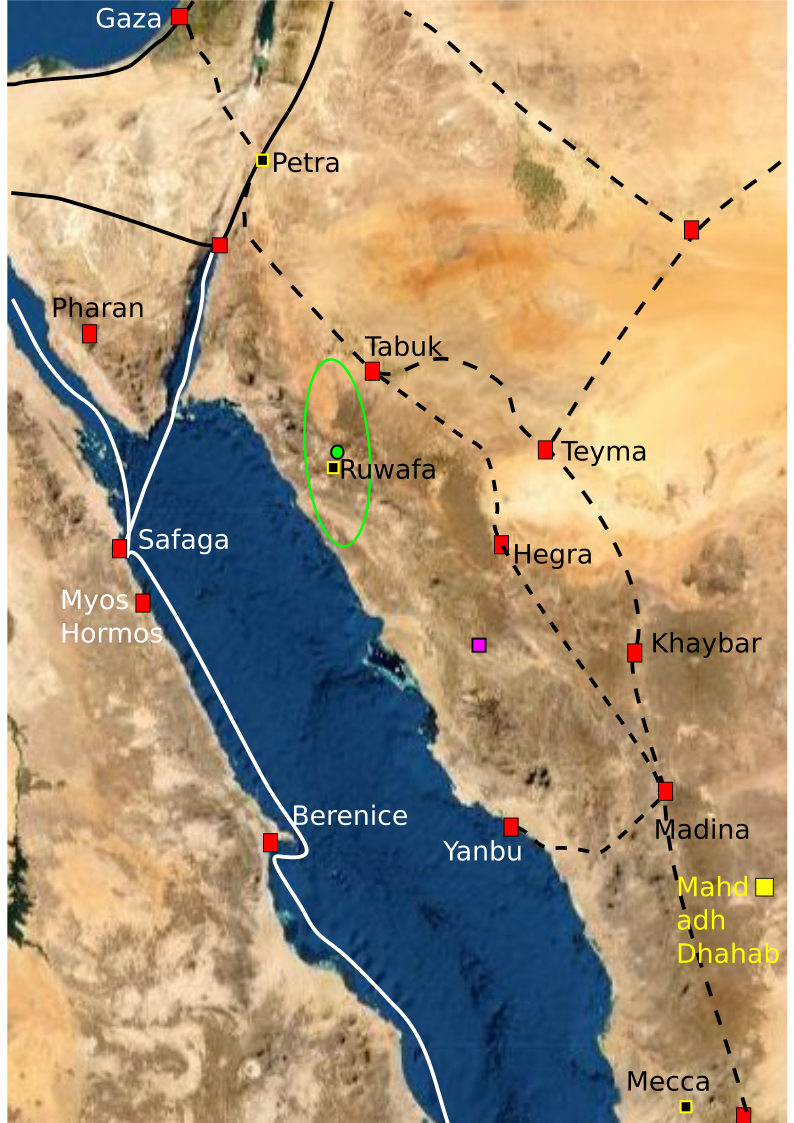}
\par\end{centering}
\caption{\label{fig:inbetweenVeds=00003D-30}The green curve is placed around
the 95 \% confidence region based on a compression with $s=-29$.
The green dot is the optimal reconstruction point of the cluster Ru.
The purple marker indicates the midpoint between Petra and Mecca along
a great circle.}
\end{figure}

In order to calculate a confidence region for the reconstruction point Ru we
fix the reconstruction points for the cluster Pe to Petra and fix
the reconstruction point of the cluster Je to the Dome of the Rock
in Jerusalem. With these reconstruction points fixed we calculate
the optimal reconstruction point of cluster Ru and calculate a confidence
region around this point. The confidence region for cluster Ru is
illustrated in Figure \ref{fig:inbetweenVeds=00003D-30}. The confidence
region does not contain any ancient settlements and it does not intersect
any of the ancient trade routes. The confidence region does not contain
the midpoint between Petra and Mecca as suggested by D. Gibson. The
optimal reconstruction point is located at $27.6664^{\circ}\,E\,36.2188^{\circ}\,N$.
The only site of archaeological relevance within the confidence region
is the Ruw\=afa Temple that lies completely isolated in the desert \cite[p. 44-49]{Macdonald2015}.
The temple has not been excavated, but some interesting inscriptions
reveal that it was built around 160 CE for Ilah (older version of Allah). The observation that the Ruw\=afa Temple
is almost at the optimal reconstruction point is new knowledge and
it is still far from clear how this result should be interpreted.

We can test Gibson's ``between qibla'' versus our ``Ruw\=afa qibla''
in the same way as we tested great circles versus rhumb lines and
the result is that ``Ruw\=afa qibla'' gives a much better description
of the data than the ``between qibla''. It is surprising that none of our reconstruction points
lie near Mecca. For instance there is an rock inscription that testify that Mashid al-Haram
was built (or rebuilt) in Mecca in 698 CE \cite[p. 111]{Amine2020}, and there are some mosques 
before the Abbaside revolution that seem to face Mecca, but they are so few that they do not lead to a separate reconstruction point.

\section*{Acknowlegment}

I thank Dan Gibson and Kristoffer Damgaard for useful comments on a draft of this paper.
\newpage{}

\bibliographystyle{IEEEtran}
\bibliography{C:/Users/peth.NB/Documents/BibTeX/database1,C:/Users/peth.NB/Documents/BibTeX/Petra}

\begin{thebibliography}{10}
\providecommand{\url}[1]{#1}
\csname url@samestyle\endcsname
\providecommand{\newblock}{\relax}
\providecommand{\bibinfo}[2]{#2}
\providecommand{\BIBentrySTDinterwordspacing}{\spaceskip=0pt\relax}
\providecommand{\BIBentryALTinterwordstretchfactor}{4}
\providecommand{\BIBentryALTinterwordspacing}{\spaceskip=\fontdimen2\font plus
\BIBentryALTinterwordstretchfactor\fontdimen3\font minus
  \fontdimen4\font\relax}
\providecommand{\BIBforeignlanguage}[2]{{%
\expandafter\ifx\csname l@#1\endcsname\relax
\typeout{** WARNING: IEEEtran.bst: No hyphenation pattern has been}%
\typeout{** loaded for the language `#1'. Using the pattern for}%
\typeout{** the default language instead.}%
\else
\language=\csname l@#1\endcsname
\fi
#2}}
\providecommand{\BIBdecl}{\relax}
\BIBdecl

\bibitem{Harremoes2008g}
\BIBentryALTinterwordspacing
P.~Harremo{\"e}s, ``Testing {G}oodness-of-{F}it via rate distortion,'' in
  \emph{Information Theory Workshop, Volos, Greece, 2009}.\hskip 1em plus 0.5em
  minus 0.4em\relax IEEE, 2009, pp. 17--21. [Online]. Available:
  \url{http://arxiv.org/abs/0903.5426}
\BIBentrySTDinterwordspacing

\bibitem{Harremoes2019}
------, ``The rate distortion test of normality,'' in \emph{2019 IEEE
  International Symposium on Information Theory (ISIT)}.\hskip 1em plus 0.5em
  minus 0.4em\relax IEEE, Jul. 2019, pp. 241--245.

\bibitem{Gibson2021}
\BIBentryALTinterwordspacing
D.~Gibson, ``Early {I}slamic qibla database,'' Jan. 2021. [Online]. Available:
  \url{https://figshare.com/articles/dataset/Early_Islamic_Qibla_Database/13570655}
\BIBentrySTDinterwordspacing

\bibitem{Sims2019}
\BIBentryALTinterwordspacing
C.~R. Sims. Ratedistortion: Routines for solving rate-distortion problems.
  Package for R program. Not updated. [Online]. Available:
  \url{https://rdrr.io/cran/RateDistortion/}
\BIBentrySTDinterwordspacing

\bibitem{Harremoes2022}
\BIBentryALTinterwordspacing
P.~Harremo{\"e}s. Arimoto{B}lahut.{R}. Package for rate distortion calculations
  in R. [Online]. Available:
  \url{http://www.harremoes.dk/Peter/ArimotoBlahut.R}
\BIBentrySTDinterwordspacing

\bibitem{Harremoes2022a}
\BIBentryALTinterwordspacing
------. Qibla{S}cript.{R}. Worksheet for analyzing qibla data in R. [Online].
  Available: \url{http://www.harremoes.dk/Peter/QiblaScript.R}
\BIBentrySTDinterwordspacing

\bibitem{Rignell1979}
K.~Rignell, \emph{Letter from {J}acob of {E}dessa to {J}ohn the {S}tylite of
  {L}itarab concerning ecclesiastical canons}.\hskip 1em plus 0.5em minus
  0.4em\relax Lund: Gleerup, 1979.

\bibitem{Hoyland2019}
R.~G. Hoyland, \emph{Seeing {I}slam As Others Saw It}.\hskip 1em plus 0.5em
  minus 0.4em\relax Gorgias Press, 2019.

\bibitem{Kalbi1952}
H.~ibn~al Kalbi, \emph{The Book of Idols}, ser. Princeton Oriental
  Studies.\hskip 1em plus 0.5em minus 0.4em\relax Princeton University Press,
  1952, vol.~14, translated by Nabih Amin Faris.

\bibitem{Grunebaum2017}
G.~E.~V. Grunebaum, \emph{Classical Islam}.\hskip 1em plus 0.5em minus
  0.4em\relax Taylor {\&} Francis Ltd., 2017.

\bibitem{Gibson2017}
D.~Gibson, \emph{Early Islamic Qiblas}.\hskip 1em plus 0.5em minus 0.4em\relax
  Vancouver: Scholars Press, 2017.

\bibitem{Crone1987}
P.~Crone, \emph{Meccan Trade and the Rise of {I}slam}.\hskip 1em plus 0.5em
  minus 0.4em\relax New Jersey: Gorgias Press, 1987.

\bibitem{Khan2021}
M.~A. Khan, \emph{The Unveiling Origin of {M}ecca}.\hskip 1em plus 0.5em minus
  0.4em\relax AuthorHouse, 2021.

\bibitem{alTamimi2019}
\BIBentryALTinterwordspacing
A.~J. al~Tamimi, ``The {B}yzantine-{A}rabic chronicle: Full translation and
  analysis.'' Aug. 2019. [Online]. Available:
  \url{http://www.aymennjawad.org/23129/the-byzantine-arabic-chronicle-full-translation}
\BIBentrySTDinterwordspacing

\bibitem{Hottinger1651}
\BIBentryALTinterwordspacing
J.~H. Hottinger, \emph{Historia orientalis, quae ex variis orientalium
  monumentis collecta.}\hskip 1em plus 0.5em minus 0.4em\relax Bodmer., 1651.
  [Online]. Available:
  \url{https://play.google.com/books/reader?id=9XJBAAAAcAAJ&pg=GBS.PA138&hl=da}
\BIBentrySTDinterwordspacing

\bibitem{Mardia2000}
K.~V. Mardia and P.~E. Jupp, \emph{Directional Statistics}, ser. Wiley Series
  in Probability and Statistics.\hskip 1em plus 0.5em minus 0.4em\relax Wiley,
  2000.

\bibitem{Harremoes2010b}
\BIBentryALTinterwordspacing
P.~Harremo{\"e}s, ``Information theory for angular data,'' in \emph{Proceedings
  Information Theory Workshop, Cairo}.\hskip 1em plus 0.5em minus 0.4em\relax
  IEEE, Jan. 2010, pp. 181--185. [Online]. Available:
  \url{www.harremoes.dk/Peter/1569245443.pdf}
\BIBentrySTDinterwordspacing

\bibitem{Blahut1972}
R.~E. Blahut, ``Computation of channel capacity and rate-distortion
  functions,'' \emph{IEEE Trans. Inform. Theory}, vol.~18, pp. 460--473, 1972.

\bibitem{Cover1991}
T.~M. Cover and J.~A. Thomas, \emph{Elements of Information Theory}.\hskip 1em
  plus 0.5em minus 0.4em\relax Wiley, 1991.

\bibitem{King2017}
D.~A. King, ``From {P}etra back to {M}akka - from "pibla" back to qibla,''
  Online on https://muslimheritage.com/pibla-back-to-qibla/, Aug. 2017.

\bibitem{King2020}
------, ``The {P}etra fallacy - early mosques do face the {S}acred {K}aaba in
  {M}ecca but {D}an {G}ibson doesn't know how / comparing historical
  orientations with modern directions can lead to false results,'' Published at
  https://muslimheritage.com/the-petra-fallacy/, Sep. 2020.

\bibitem{Gibson2021a}
\BIBentryALTinterwordspacing
D.~Gibson, ``Navigation and the qibla: 5: The {I}ndian circle,'' Youtube video,
  Jan. 2021. [Online]. Available:
  \url{https://www.youtube.com/watch?v=hEm65krB6W8&list=PLneIIflWyUahE4fg25U9jr1km1gXusDtm&index=35&t=6s&ab_channel=DanGibson}
\BIBentrySTDinterwordspacing

\bibitem{Gibson2018}
\BIBentryALTinterwordspacing
------, ``Sidi {G}hanem {M}osque,'' 2018. [Online]. Available:
  \url{https://nabataea.net/explore/cities_and_sites/sidi-ghanem-mosque/}
\BIBentrySTDinterwordspacing

\bibitem{Voss2014}
J.~Voss, \emph{An Introduction to Statistical Computing}.\hskip 1em plus 0.5em
  minus 0.4em\relax Wiley, 2014.

\bibitem{Macdonald2015}
M.~C.~A. Macdonald, \emph{Arabs and Empires before {I}slam}.\hskip 1em plus
  0.5em minus 0.4em\relax Oxford University Press, 2015, ch. Arabs and Empires
  before the Sixth Century, pp. 11--89.

\bibitem{Amine2020}
A.~Amine, \emph{L'islam de P\'etra R\'eponse \`a la th\`ese de Dan Gibson:
  Pr\'esentation \& Revue critique}.\hskip 1em plus 0.5em minus 0.4em\relax BoD
  - Books on Demand, 2020.

\end{thebibliography}

\end{document}